\documentclass{article}
\PassOptionsToPackage{numbers,sort&compress}{natbib}
\usepackage[preprint]{neurips_2026}
\usepackage[T1]{fontenc}
\usepackage{amsmath,amssymb,booktabs,longtable,tabularx,graphicx,xcolor}
\usepackage{microtype,float,listings}
\usepackage[colorlinks=true,linkcolor=blue!50!black,citecolor=blue!50!black,urlcolor=blue!50!black]{hyperref}

\newcommand{\captionbelowtable}{\normalsize\setlength{\abovecaptionskip}{7pt}\setlength{\belowcaptionskip}{0pt}}

\title{JEV-Star: Fast, Low-Cost StarCraft II Control\\with Language-Model Planning}
\author{%
  Weiyu Ma \quad Liangbing Zhao \quad Yongcheng Zeng$^{1}$ \quad Jian Zhao$^{2}$\\
  $^{1}$Institute of Automation, Chinese Academy of Sciences\\
  $^{2}$Beijing Zhongguancun Academy\\
  \texttt{sc2meisah@gmail.com} \quad \texttt{jianzhao@zgci.ac.cn}
}

\begin{document}
\maketitle

\begin{abstract}
We present JEV-Star, a StarCraft II controller that defeats the strongest non-cheating built-in AI, Lv7, by combining fast JEV action selection with persistent GPT-6 planning. The combined system wins four full games at Lv5--Lv7, including two Lv7 victories with different seeds, while retaining a median JEV response time of 0.422 seconds. Mean estimated model cost across these games is USD~3.71 per game: USD~0.15 for JEV and USD~3.56 for GPT-6. We compare this system with an initial JEV-only controller in full-game macro control and multi-unit micromanagement. The standalone controller reaches a 20-minute limit against Lv2 without expanding. Across 35 battle maps with three episodes per map and controller, the combined system raises mean enemy elimination from 16.50\% to 37.69\% and wins from 3 to 7 out of 105. Replay frames and decision logs document resource reservation, persistent economic goals, and stable army objectives in successful full games. The results demonstrate a practical division between inexpensive, subsecond decisions and longer-horizon planning. The comparison evaluates complete systems; concurrent interface improvements mean that planning's contribution is not isolated by a controlled ablation. Code is available at \url{https://github.com/sc2musa/Jev_Star}
\end{abstract}

\section{Introduction}

StarCraft II requires both immediate decisions and commitments whose benefits arrive much later. A controller can repeatedly choose legal, affordable actions yet never save enough to expand its economy. It can also alternate between attacking and retreating without carrying either decision through. These failures motivate a separation between choosing the next action and maintaining a strategy.

We implement this separation with JEV, a hosted model that selects among structured alternatives \citep{typesafe2026jev}, and GPT-6, which supplies persistent strategic guidance. JEV remains responsible for frequent action selection. GPT-6 specifies goals, spending priorities, and conditions for changing army posture. The intended benefit is to preserve fast, low-cost interaction while using general-purpose planning where a single-step decision is insufficient.

JEV-Star builds on LLM Play SC2 for full-game control and SMAC-Hard for micromanagement \citep{ma2023textsc2,deng2024smachard}. We compare JEV-only with JEV + GPT-6. The combined controller achieves two victories against Lv7, the highest non-cheating difficulty of StarCraft II's built-in AI \citep{blizzard2026sc2protocol}. Our experiments measure game outcomes, response latency, and the combined model cost of a complete game. Replay frames and decision logs show how persistent plans change economic development and army behavior while preserving a fast action loop.

\section{Related work}

StarCraft II AI spans the SC2LE interface, full-game agents such as TStarBots, AlphaStar, TStarBot-X, and StarCraft Commander, and offline learning with AlphaStar Unplugged \citep{vinyals2017sc2le,sun2018tstarbots,vinyals2019alphastar,han2020tstarbotx,wang2020scc,mathieu2023unplugged}. Cooperative combat is studied in SMAC and its extensions, with methods including QMIX and MAPPO \citep{samvelyan2019smac,ellis2022smacv2,deng2024smachard,rashid2018qmix,yu2021mappo}. Our contribution concerns inference-time use of hosted models, without task-specific weight training.

LLM Play SC2 and LLM-PySC2 establish language-model interfaces for game control \citep{ma2023textsc2,li2024llmpysc2}. ReAct, Reflexion, Voyager, and SayCan connect reasoning or persistent context to executable actions \citep{yao2022react,shinn2023reflexion,wang2023voyager,ahn2022saycan}. AgentBench and WebArena emphasize evaluating interactive outcomes \citep{liu2023agentbench,zhou2023webarena}. JEV-Star uses a native choice interface for concrete actions and retains an explicit plan across decisions.

FrugalGPT and RouteLLM reduce cost through model selection, while ReWOO and LLMCompiler organize reasoning and tool execution \citep{chen2023frugalgpt,ong2024routellm,xu2023rewoo,kim2023llmcompiler}. We assign models by control role: frequent JEV actions and occasional GPT-6 plans. Our experiments examine this assignment through game performance, response latency, and per-game expenditure.

\section{Method}

\subsection{Structured action selection with JEV}

\begin{figure}[t]
\centering
\includegraphics[width=\linewidth]{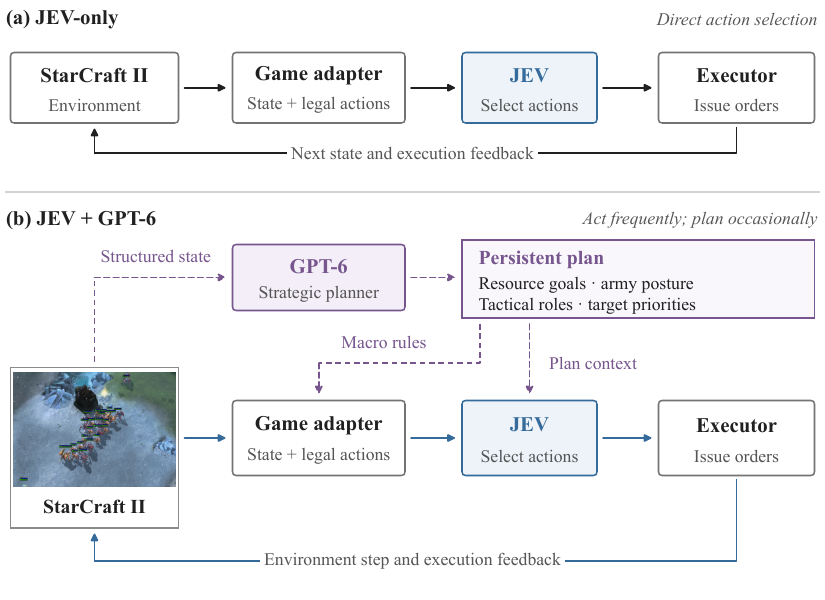}
\caption{Action selection and persistent planning in JEV-Star. (a) JEV-only selects legal actions from the current state and recent feedback. (b) JEV + GPT-6 adds an occasional planning path (dashed purple arrows) to the frequent action loop (solid blue arrows). The retained plan supplies context to JEV and constrains macro candidates through resource reservations and army-posture rules. Micro plans provide tactical context while retaining legal unit choices. Macro and micro use separate adapters. The replay thumbnail illustrates the environment; models receive structured observations.}
\label{fig:architecture}
\end{figure}

At decision time \(t\), the adapter constructs a structured observation \(x_t\) and an executable candidate set \(\mathcal{A}_t\). JEV selects an identifier:
\begin{equation}
    a_t = \operatorname{JEV}(x_t,\mathcal{A}_t).
\end{equation}
The executor validates the identifier, attempts the associated command, and includes recent outcomes in subsequent observations. Both methods use this loop. The combined controller adds an explicit plan (Figure~\ref{fig:architecture}); the evaluated systems also differ in candidate descriptions and execution handling, as detailed in the experimental setup.

For macro control, the observation summarizes resources, workers, buildings, technology, army composition, visible enemies, and recent actions. Candidates represent high-level commands such as producing a unit, constructing a building, researching an upgrade, expanding, attacking, retreating, or waiting. Local game code handles placement and low-level order execution. Feasibility checks consider prerequisites, supply, resources, and available producers. JEV chooses one command per request.

For micromanagement, the observation describes the team and visible enemies, including health, positions, ranges, cooldowns, and energy. Each living unit receives its own legal alternatives: move to an enumerated destination, attack an available target, heal an ally where supported, or stop. Unit choices share the same team observation and are submitted in concurrent batches. Execution returns information about rejected actions. This is centralized shared-vision control; it does not implement independently observed, decentralized MARL policies.

The JEV-only method selects directly from current state and short execution history. It has no separate process that maintains a future resource budget or a persistent team objective. In particular, an affordability filter can hide an expansion until sufficient minerals are available, even though reaching that threshold requires declining currently affordable purchases.

\subsection{Persistent planning with GPT-6}

In the combined method, a planner reads a state summary and, when available, the previous plan. Let \(s_k\) denote the observation index used for planning and \(t_k\geq s_k\) the first decision that can use the validated result. The guidance \(z_k\) remains active across multiple action decisions:
\begin{equation}
    z_k=\operatorname{GPT\text{-}6}(x_{s_k},z_{k-1}), \qquad
    a_t=\operatorname{JEV}(x_t,z_k,\widetilde{\mathcal{A}}_t), \quad
    t_k\leq t<t_{k+1}.
\end{equation}
The planner proposes goals rather than individual engine commands. Its output is checked before becoming active. A failed or delayed planning request leaves the previous valid plan available.

A macro plan specifies an economic objective, worker and base targets, production priorities, resource reservations, and an army posture. Resource reservation connects a future purchase to present action availability: when saving for an expansion, the adapter can suppress lower-priority spending until the required minerals accumulate. Emergency supply or defense actions can override these restrictions. Army guidance includes separate thresholds for starting and ending an attack and a minimum commitment interval, reducing repeated reversals near a single threshold. The plan enters both JEV's context and the adapter's candidate rules; the latter makes strategic commitments operational.

A micro plan describes conditional tactics and roles, such as which units should absorb pressure, which targets deserve priority, and when ranged units should reposition. One plan is generated for each map and reused for its three episodes. The legal action set remains available to JEV. A common plan provides shared intent, but independently chosen unit actions still lack a joint reservation mechanism for damage or healing targets.

\subsection{Control timing}

Full-game control is asynchronous: the game continues while requests are in flight. JEV requests are rate-limited to at most one per wall-clock second. The planner is checked on a nominal 60-game-second interval and at relevant events, subject to a minimum execution window. Completed plans are applied to subsequent decisions.

Micromanagement uses fixed-step interaction. The environment advances eight game loops only after a complete team decision has returned. Concurrent batching reduces waiting compared with sequential unit requests, but the slowest required response determines the next step. Consequently, model response latency and whole-episode wall time measure different parts of this system.

Neither method fine-tunes JEV or GPT-6. Both rely on the same division of responsibilities: game code determines observations and executable commands, JEV chooses actions, and the combined method adds GPT-6's persistent goals.

\section{Experiments}

\subsection{Setup and measurements}

We compare the initial JEV-only controller with the completed JEV + GPT-6 system. The runs use JEV 1.13, GPT-6 Astra with medium reasoning effort, and StarCraft II 5.0.16. Full-game experiments play Protoss against the built-in Zerg opponent on AltitudeLE, with eight starting workers. We report difficulty as Lv2, Lv5, Lv6, Lv7, or Lv8, following the ordered levels used by LLM Play SC2 \citep{ma2023textsc2}. The main sample contains one JEV-only game and four combined-system games; a completed Lv8 stress test is reported separately.

Micromanagement covers 35 local SMAC-Hard maps, with three episodes per map and method: 105 episodes each. Both methods use the same map configurations, game version, opponent controls, shared-vision setting, and disabled optional special abilities. Two local development maps are also excluded in a separate 33-map summary. There is one shared plan per map, so the three combined-method episodes do not represent independent planner samples. Candidate descriptions and execution handling also improved between the two controllers; these experiments measure the implemented systems rather than an isolated planner ablation.

We report wins, draws, losses, and mean enemy elimination, defined per episode as the fraction of initial enemy units eliminated. Each episode receives equal weight. For efficiency, we measure recorded response latency, total wall time, and model cost per complete episode. Macro outcomes and game durations are checked against replays. Cost estimates include recorded calls with known token usage, whether or not a returned plan was accepted.

\subsection{Full-game performance: defeating the strongest non-cheating AI}

Table~\ref{tab:macro} shows that JEV + GPT-6 wins all four completed non-cheating matches, including two victories against Lv7, the strongest built-in opponent without cheating advantages. These wins take 10:49--15:55 of game time. The first attacks occur between 7:22 and 8:11, following an economic and production buildup. JEV-only reaches its 20-minute game-time limit without a win. The two Lv7 seeds establish successful play at the highest non-cheating difficulty, although the sample is too small to estimate a reliable win rate.

\begin{table}[t]
\centering\small
\begin{tabular}{llclrr}
\toprule
Method & Difficulty & Seed & Outcome & Game time & Wall time \\
\midrule
JEV-only & Lv2 & 1 & Time limit & 20:00 & 20:40 \\
JEV + GPT-6 & Lv5 & 1 & Win & 13:28 & 14:00 \\
JEV + GPT-6 & Lv6 & 1 & Win & 10:49 & 11:18 \\
JEV + GPT-6 & Lv7 & 1 & Win & 12:48 & 13:26 \\
JEV + GPT-6 & Lv7 & 2 & Win & 15:55 & 16:31 \\
JEV + GPT-6 & Lv8 & 1 & Loss & 10:14 & 10:43 \\
 
\bottomrule
\end{tabular}
\captionbelowtable
\caption{Completed full-game experiments (times in minutes:seconds). The first five rows form the main comparison. The final Lv8 row is a separate stress test against an opponent with an information advantage. A time limit is not a victory.}
\label{tab:macro}
\end{table}

\begin{figure}[tp]
\centering
\includegraphics[width=\linewidth]{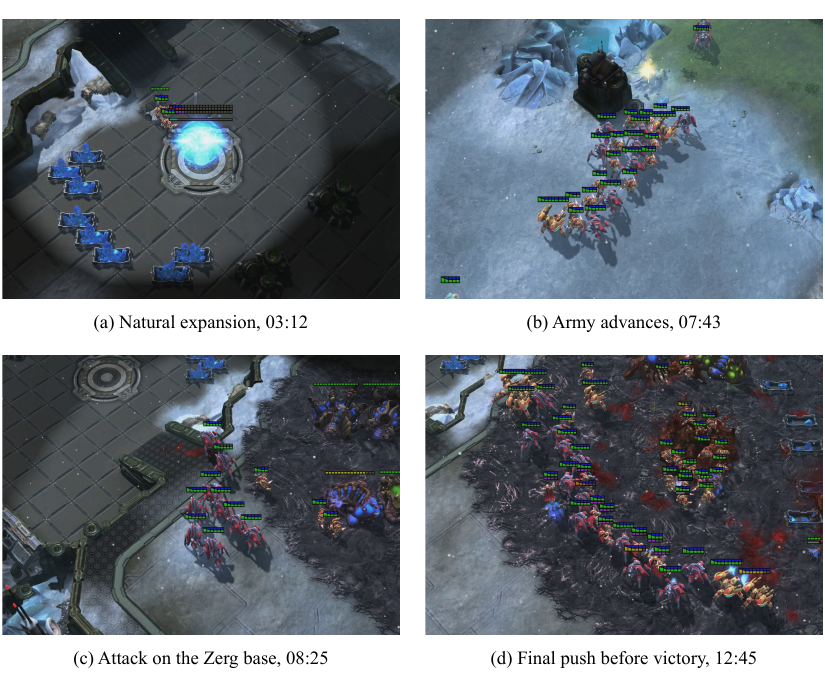}
\caption{Actual replay frames from JEV + GPT-6 defeating Lv7 on AltitudeLE, seed 1. (a) The natural Nexus is under construction. (b) The army advances after the logged attack command at 07:33. (c) Units attack the Zerg base. (d) The push continues shortly before victory at 12:48. Displayed times are game time, rounded to the nearest second. The replay observer view is used only for presentation.}
\label{fig:lv7replay}
\end{figure}

Figure~\ref{fig:lv7replay} shows four moments from the Lv7 seed-1 victory: expansion, army movement, pressure on a Zerg base, and the final push. The frames come from the evaluated replay, rendered afterward with an observer camera. This viewing perspective supplies illustrations only; the controller acts on structured observations, not these images.

LLM Play SC2 reports its main evaluation at Lv5 (12 wins in 20 games for GPT-4-Turbo); a separate GPT-3.5 prompt study reaches Lv6 with an 8.33\% win rate \citep{ma2023textsc2}. Our Lv7 victories exceed the demonstrated difficulty in those reported experiments. The game versions and starting conditions differ, so this is a difficulty milestone rather than a matched win-rate comparison. That paper does not provide directly comparable per-game model charges or response latencies; we therefore quantify JEV-Star's speed and expenditure from our own logs in Section~\ref{sec:efficiency}.

\paragraph{Stress test with privileged vision.}
We additionally test Lv8, whose built-in opponent has a vision advantage. This game ends in defeat at 10:14. At five minutes, the agent's army resource value is 900 versus the opponent's 1,525; its first Immortal appears only at 6:25. It never issues an attack command. This failure illustrates a remaining economic--military imbalance when facing a stronger, privileged opponent; it is separate from the non-cheating difficulty milestone.

\subsection{Full-game decisions: build order, technology, and unit allocation}
\label{sec:macrochoices}

\begin{figure}[t]
\centering
\includegraphics[width=\linewidth]{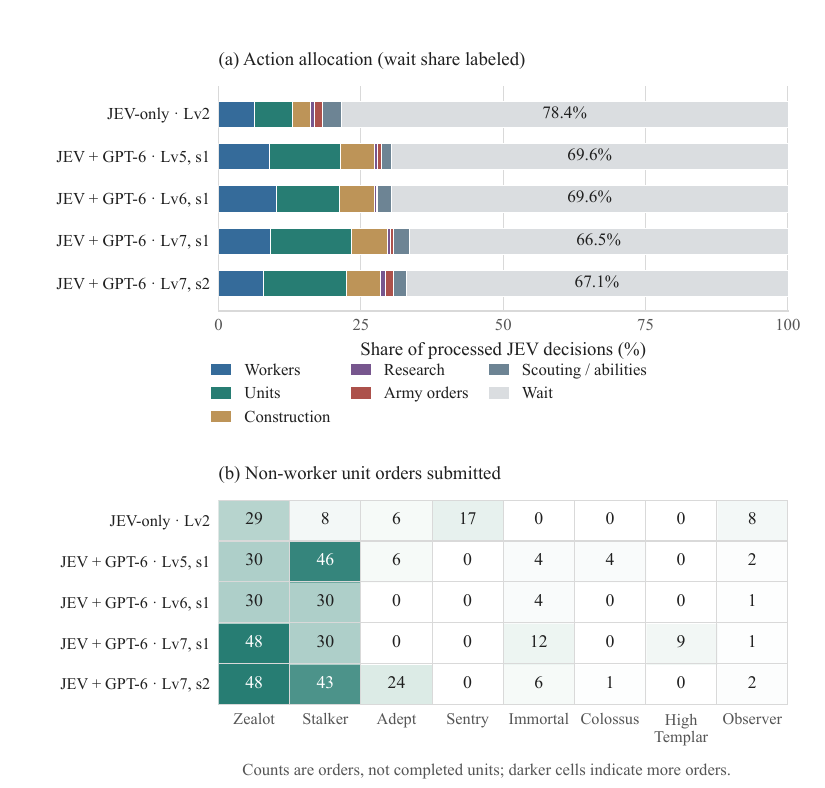}
\caption{Full-game JEV decisions and production. (a) Action categories across all processed model choices in each main game, including explicit waits; gray segments label the wait percentage. (b) Submitted production orders for every non-worker unit type produced in these games. Counts are model-directed engine orders, not completed units or surviving army composition. No air combat unit is ordered in any of these five games. Candidate availability and game duration differ across runs; unit counts therefore describe the implemented controllers' allocation, not an intrinsic model preference.}
\label{fig:macrochoices}
\end{figure}

\paragraph{What does JEV choose?}
Figure~\ref{fig:macrochoices} accounts for every JEV decision forwarded to the executor in the five main games. The standalone run contains 1,020 such decisions; the combined runs contain 737, 583, 701, and 850. Each action event is joined to its original model response and candidate set. We count an explicit wait as a decision, and separate model choices from the engine orders they submit. Local worker distribution, automatic Gateway morphing, and continued execution of army intent are outside this count. Responses discarded before execution are also excluded.

JEV-only chooses to wait in 800 of 1,020 decisions (78.4\%); the combined controller waits in 1,955 of 2,871 (68.1\%). Non-worker production accounts for 6.7\% versus 13.3\% of processed choices. The remaining decisions allocate attention to workers, construction, research, scouting, abilities, and army commands. A wait is not necessarily a mistake: saving resources, occupied producers, and ongoing army orders can all justify it. These frequencies describe complete games of different lengths and difficulties, rather than a matched measure of action quality.

\paragraph{From technology purchases to usable forces.}
The standalone controller submits 29 Zealot, 17 Sentry, eight Stalker, six Adept, and eight Observer production orders, but no Immortal orders. The combined runs instead submit 156 Zealot, 149 Stalker, 30 Adept, 26 Immortal, five Colossus, nine High Templar, and six Observer orders in total. Each of the four wins includes Immortals. This is a shift toward a ground army with sustained basic-unit production and higher-tier support, rather than simply a larger variety of purchases.

Availability matters when interpreting these choices. In the standalone run, Immortal production is offered at only three processed decisions and never selected; the combined runs select it in 26 of 34 opportunities. Sentries appear in 58 standalone candidate sets and are chosen 17 times, but never appear in the combined candidate sets. Thus, the absence of combined-system Sentries cannot be attributed to JEV rejecting them. The plan, spending rules, available producers, and model selection jointly determine the observed unit allocation. Appendix~\ref{app:macrochoices} reports these counts explicitly.

\paragraph{Opening order and technology progression.}
Table~\ref{tab:technology} gives the first submitted construction, research, and production orders. All four combined games follow Gateway, Cybernetics Core, early expansion, and Robotics development. Their Core orders occur at 2:01--2:09 and Robotics Facility orders at 3:31--4:01, compared with 2:41 and 4:50 for JEV-only. Warp Gate research starts at 2:42--2:52 in the combined runs versus 3:20 for JEV-only; its first observed completion is at 4:18--4:24 versus 5:00. These are distinct milestones: ordering research does not make it immediately available.

The standalone controller constructs a Forge at 1:41, before its Core, and later builds a Stargate and completes air-weapons and air-armor upgrades without ordering an air combat unit. Its Robotics Facility produces only Observers. These investments broaden the technology tree without yielding an expansion or higher-tier combat force. The combined games prioritize expansion and Robotics before adding optional branches: two reach Colossus production, while one Lv7 game adds High Templars. The latter still initiates no Psionic Storm research, illustrating that a winning trajectory need not use every technology investment well.

\begin{table}[t]
\centering\small
\setlength{\tabcolsep}{5pt}
\begin{tabular}{lrrrrr}
\toprule
& JEV-only & \multicolumn{4}{c}{JEV + GPT-6} \\
\cmidrule(lr){3-6}
First submitted order & Lv2 & Lv5, s1 & Lv6, s1 & Lv7, s1 & Lv7, s2 \\
\midrule
Gateway & 01:06 & 01:13 & 01:16 & 01:11 & 01:08 \\
Cybernetics Core & 02:41 & 02:03 & 02:05 & 02:01 & 02:09 \\
Warp Gate research & 03:20 & 02:48 & 02:47 & 02:42 & 02:52 \\
Expansion Nexus & -- & 02:42 & 02:39 & 02:57 & 02:46 \\
Robotics Facility & 04:50 & 03:35 & 03:43 & 04:01 & 03:31 \\
First Immortal order & -- & 05:44 & 05:30 & 05:38 & 05:48 \\
Twilight Council & 03:49 & -- & 08:40 & 06:44 & 07:55 \\
Charge & 04:38 & -- & 09:22 & 07:24 & 10:04 \\
Robotics Bay & -- & 08:47 & -- & -- & 09:59 \\
First Colossus order & -- & 09:51 & -- & -- & 11:28 \\
Templar Archives & -- & -- & -- & 09:51 & -- \\
First High Templar order & -- & -- & -- & 10:34 & -- \\
Forge & 01:41 & 11:49 & 10:15 & 10:58 & 12:07 \\
Stargate & 05:45 & -- & -- & -- & -- \\
Air weapons 1 & 05:04 & -- & -- & -- & -- \\
Air armor 1 & 11:16 & -- & -- & -- & -- \\
 
\bottomrule
\end{tabular}
\captionbelowtable
\caption{Technology and production milestones in game time (minutes:seconds), rounded to the nearest second. Each entry is the first order with a positive submitted-order count; a dash means no such order in that game. Times are submission times, not building completion, upgrade completion, or unit birth. Nexus refers to the first expansion beyond the starting base.}
\label{tab:technology}
\end{table}

The complete opening construction sequence for JEV-only and the Lv7 seed-1 win appears in Appendix~\ref{app:macrochoices}. Their first Pylon, gas, and Gateway timings are similar; the subsequent investment order diverges. The Lv7 controller builds its Core before expanding at 2:57, then adds another Gateway and Assimilator. The expansion order uses an observation with 22 workers, although the plan suggests expanding around 19. This gap distinguishes strategic guidance from the executed build order and shows why both must be inspected.

\subsection{Micromanagement: greater enemy elimination, limited wins}

Table~\ref{tab:micro} summarizes the complete 35-map comparison. Mean enemy elimination rises from 16.50\% to 37.69\%, an increase of 21.19 percentage points, or 2.28 times the initial value. Wins increase from 3 to 7 out of 105. Excluding the two development maps retains the same win counts and increases mean elimination from 15.55\% to 39.17\%. The improvement is therefore not driven by those maps.

\begin{table}[t]
\centering\small
\begin{tabular}{lrrrr}
\toprule
& \multicolumn{2}{c}{All 35 maps: 105 episodes} & \multicolumn{2}{c}{33 maps: 99 episodes} \\
\cmidrule(lr){2-3}\cmidrule(lr){4-5}
Method & W/D/L & Elimination & Wins & Elimination \\
\midrule
JEV-only & 3/2/100 & 16.50\% & 3/99 & 15.55\% \\
JEV + GPT-6 & 7/0/98 & 37.69\% & 7/99 & 39.17\% \\
 
\bottomrule
\end{tabular}
\captionbelowtable
\caption{Micromanagement outcomes. Elimination is the episode-average fraction of enemy units eliminated. The 33-map subset excludes two local development maps. Both methods use three episodes per map.}
\label{tab:micro}
\end{table}

\begin{figure}[t]
\centering
\includegraphics[width=\linewidth]{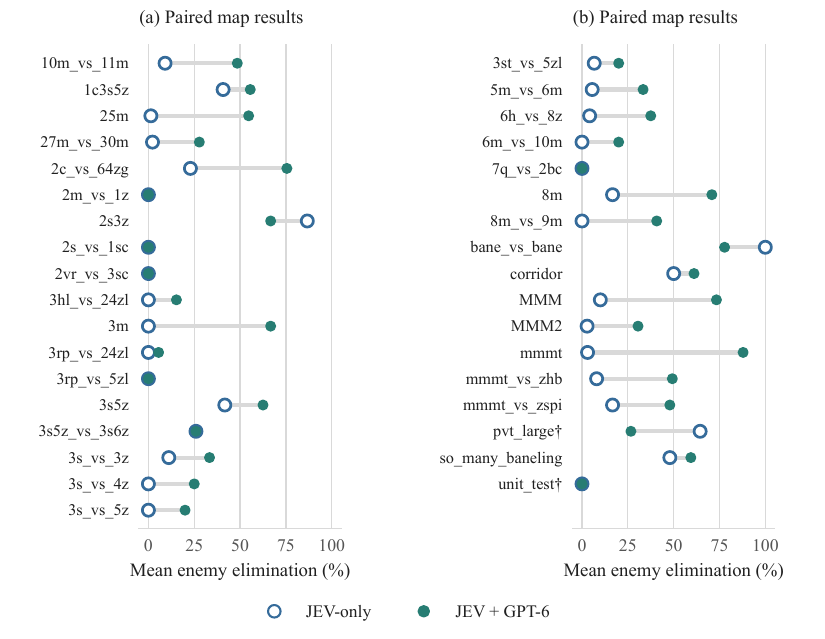}
\caption{Mean enemy elimination on all 35 maps, with three episodes per point. A connecting line pairs the two methods on the same map. The combined system improves 25 maps, ties seven, and declines on three. Daggers mark the two local development maps excluded from the 33-map summary.}
\label{fig:micro}
\end{figure}

The combined system improves mean elimination on 25 maps, ties on seven, and declines on three (Figure~\ref{fig:micro}). On \texttt{mmmt}, elimination increases from 3.03\% to 87.88\% and wins from zero to two. On \texttt{MMM}, it rises from 10.00\% to 73.33\%, with one win; \texttt{8m} and \texttt{2c\_vs\_64zg} also acquire wins. These cases indicate a useful change in tactical behavior beyond merely issuing valid actions.

Nevertheless, the combined system still loses 98 episodes. On \texttt{3m}, elimination increases from zero to 66.67\% without a win. On \texttt{2s3z}, wins decline from two to one; on \texttt{bane\_vs\_bane}, elimination declines from 100\% to 77.78\%. The combined controller improves many battles but does not consistently convert enemy elimination into team survival and victory.

\begin{figure}[t]
\centering
\includegraphics[width=\linewidth]{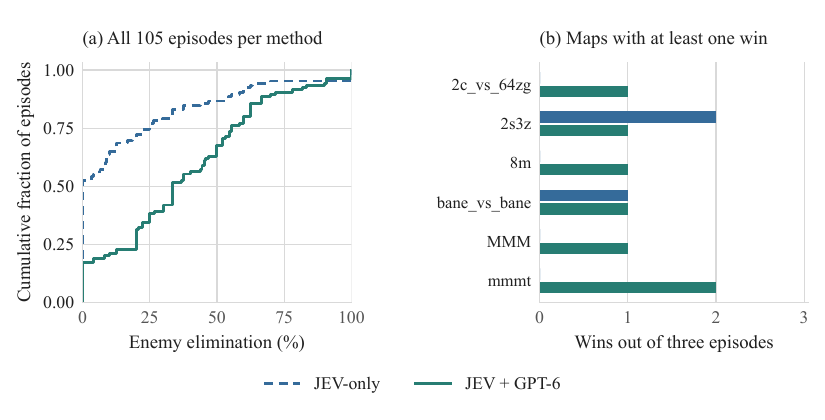}
\caption{Distribution of micromanagement outcomes. (a) Empirical cumulative distributions over all 105 episodes per method; a lower curve at an elimination threshold indicates fewer episodes at or below that threshold. (b) Wins on every map where either method wins at least once; both methods have zero wins on the remaining 29 maps. Each map has three episodes per method. These are descriptive distributions; the three combined-method episodes share a map-specific plan.}
\label{fig:outcomes}
\end{figure}

Figure~\ref{fig:outcomes} shows the distribution behind these averages. Episodes with no enemies eliminated fall from 55 to 18, while episodes eliminating at least half the enemy units increase from 14 to 39. Victories remain concentrated: JEV-only wins on two maps and JEV + GPT-6 on six. Elimination and victory are distinct outcomes; a simultaneous wipeout can be a draw, while eliminating all attacking opponents can yield a win with enemy support units still alive.

\begin{figure}[tp]
\centering
\includegraphics[width=\linewidth]{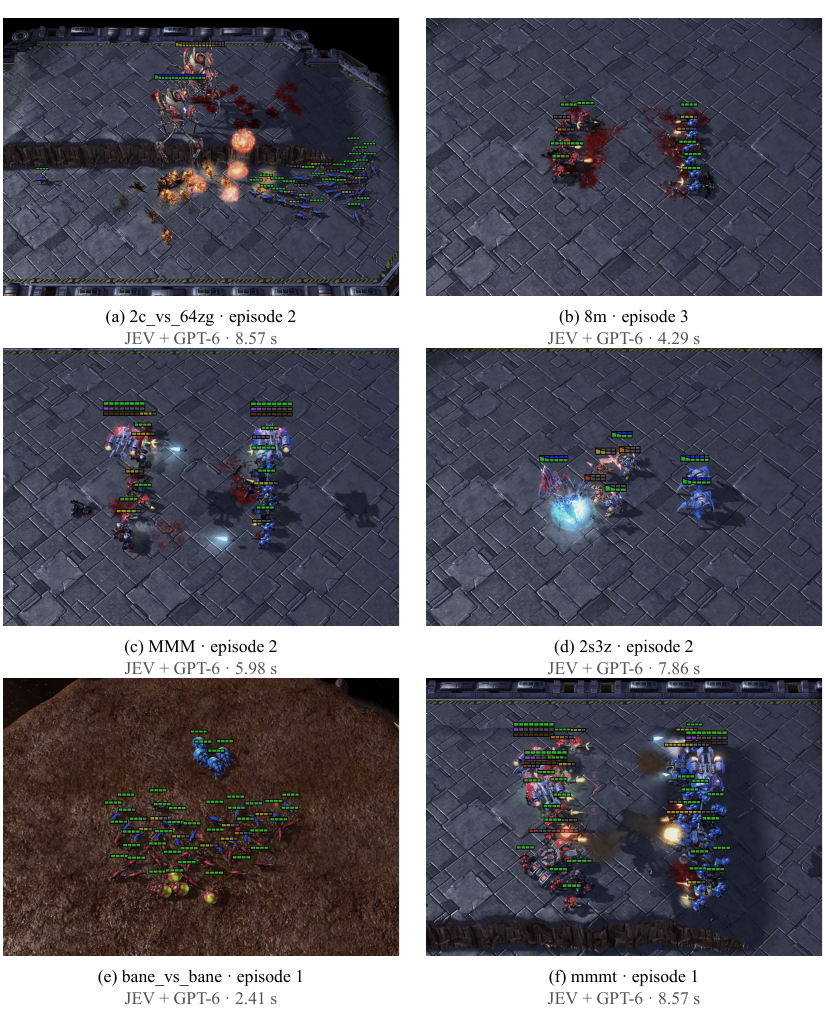}
\caption{In-game views of all six maps on which a controller wins. Each panel shows combat at 40\% of the first winning JEV + GPT-6 episode on that map, with map name, episode number, and elapsed episode time below the image. These are decoded SC2 replay frames, not generated illustrations. Allied units are red and enemies blue. The full 105-episode comparison determines performance; these panels document the situations behind the wins. Engagement and terminal views of every recorded winning episode appear in Appendix~\ref{app:allwins}.}
\label{fig:winningmaps}
\end{figure}

Figure~\ref{fig:winningmaps} shows real in-game combat on all six maps with recorded victories: \texttt{2c\_vs\_64zg}, \texttt{8m}, \texttt{MMM}, \texttt{2s3z}, \texttt{bane\_vs\_bane}, and \texttt{mmmt}. Each panel comes from the first winning combined-system episode on that map. The examples include asymmetric Colossus--Zergling combat, infantry battles, mixed ground forces, and explosive engagements. Appendix~\ref{app:allwins} shows engagement and terminal frames for all ten recorded wins: three JEV-only and seven combined-system episodes. Outcomes come from the episode summaries, and all displayed clocks restart at the corresponding episode boundary.

Appendix~\ref{app:replays} also retains the paired first \texttt{mmmt} episodes, including the JEV-only defeat, alongside tactical guidance and decision traces. These images and the victory galleries illustrate existing evaluation episodes; they do not increase the sample size.

\subsection{Why does planning help?}
\label{sec:behavior}

\paragraph{Saving for a future goal.}
The initial JEV-only full game illustrates a failure of longer-horizon resource allocation. It grows to 68 workers but remains on one base for the entire 20 minutes. Across 1,024 requests, the largest observed mineral balance is 335, below the 400 required for a Nexus. Expansion is consequently absent from every candidate set. Repeated affordable purchases prevent the saving needed to make the strategic action available.

\begin{figure}[t]
\centering
\includegraphics[width=\linewidth]{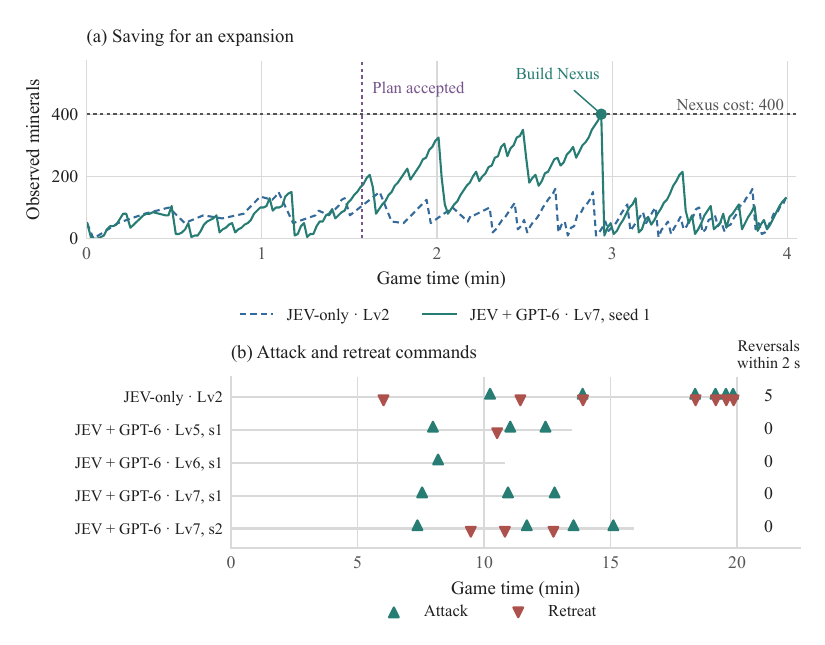}
\caption{Recorded economic and army commitments. (a) Mineral observations from JEV requests during the first four game minutes: JEV-only at Lv2 and JEV + GPT-6 at Lv7, seed 1. The combined controller accepts an expansion plan at 1:34 and submits a Nexus order at 2:57. The horizontal line marks the 400-mineral requirement. (b) Attack and retreat commands across the five main full games; the right column counts attack-to-retreat reversals within two game seconds. Horizontal segments end at each game's termination. These examples use different opponents and compare complete controllers.}
\label{fig:commitments}
\end{figure}

\begin{figure}[t]
\centering
\includegraphics[width=\linewidth]{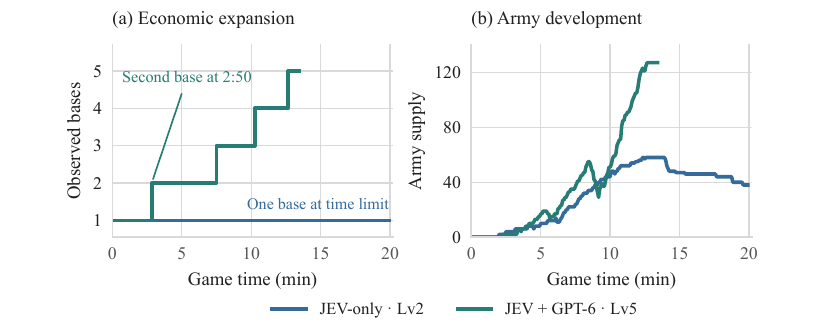}
\caption{Example full-game trajectories: JEV-only at Lv2 and JEV + GPT-6 at Lv5. The standalone controller remains on one base; the combined controller saves for expansion and develops a larger army. Base counts include structures under construction, and army supply can include pending production. The different opponents make this a behavioral illustration, not a controlled ablation.}
\label{fig:planning}
\end{figure}

In the Lv7 seed-1 game, the first accepted plan requests two bases and prioritizes the natural expansion around 19 workers. The adapter implements the expansion priority through a spending reservation. Figure~\ref{fig:commitments}(a) shows the request-time mineral balance reaching 400 before JEV selects the build action. Across the four wins, a second Nexus is first observed at 2:50--3:05, and peak observed base counts reach four to six. Figure~\ref{fig:planning} contrasts the economic and army trajectories. The combined controller thus couples an explicit future target with a mechanism that protects the resources needed to reach it.

This evidence diagnoses the initial controller's planning limitation, rather than showing that JEV can never reason about expansion. The affordability filter and one-step objective interact: neither exposes the value of temporarily refusing purchases. The combined system supplies that missing context and commitment.

Table~\ref{tab:replaytrace} follows the same Lv7 game shown in Figure~\ref{fig:lv7replay}. The log connects a request to prioritize the natural expansion, JEV's choice when 400 minerals become available, and the subsequent appearance of a second Nexus. A later attack plan is followed by 24 submitted unit attack orders within one game second. This trace distinguishes model intent, command submission, and observed game state. Selected original fields appear in Appendix~\ref{app:replays}.

\begin{table}[t]
\centering\small
\setlength{\tabcolsep}{4pt}
\begin{tabularx}{\linewidth}{llX}
\toprule
Game time & Source & Recorded behavior \\
\midrule
01:34.286 & GPT-6 plan & Start the natural around 19 workers before filling army ceilings. \\
02:56.250 & JEV observation & 400 minerals; one Nexus; expansion is an available choice. \\
02:56.786 & JEV and executor & Select BUILD NEXUS in 391 ms; submit one build order. \\
03:05.179 & Game state & A second Nexus is observed, including construction. \\
07:32.321 & GPT-6 plan & Attack posture; 44 ready army supply; 30-second commitment. \\
07:33.214 & Executor & Submit 24 attack orders under the new plan. \\
12:47.679 & Game result & Victory. \\
 
\bottomrule
\end{tabularx}
\captionbelowtable
\caption{Plan--decision--execution trace from the Lv7 seed-1 victory. Rows summarize contemporaneous logs; time is minutes:seconds.milliseconds. A submitted building order is followed by an independent observation of construction.}
\label{tab:replaytrace}
\end{table}

\paragraph{Maintaining an army objective.}
In the same JEV-only game, six attack commands and seven retreat commands include five attack-to-retreat reversals within two game seconds. Such rapid reversals leave little time for an attack to make progress. None of the four combined-system wins contains a reversal within that interval (Figure~\ref{fig:commitments}(b)). The combined controller uses persistent posture, separate attack and retreat thresholds, and a commitment window to maintain the objective across JEV calls. Peak observed army supply rises from 58 in the standalone run to 79--129 in the combined games. Different opponents and other controller changes preclude a causal estimate, but the traces are consistent with better strategic continuity.

\paragraph{Shared intent does not resolve all coordination.}
Micromanagement reveals a different boundary. In the combined system, 2,629 of 2,662 unit decisions with a ready weapon and an available in-range target choose an in-range attack. Basic engagement is usually sensible. However, among 930 decisions with a long weapon cooldown, nearby melee pressure, and an available movement endpoint that increases separation, 794 continue attacking and only 63 choose such a separating endpoint. The controller often fails to use movement during downtime.

Healing exposes resource conflicts between units. Execution records contain 500 healing rejections because a target is already targeted and ten for insufficient energy. Among 378 subsequent decisions that expose a previously rejected healing target together with an alternative legal target, 243 repeat the previous target. These conditional counts identify recurring behavior, not optimal-action labels. They suggest that planning alone cannot replace precise cooldown control, updated execution feedback, or joint allocation of healing targets.

\subsection{Response time and cost per game}
\label{sec:efficiency}

\begin{figure}[t]
\centering
\includegraphics[width=\linewidth]{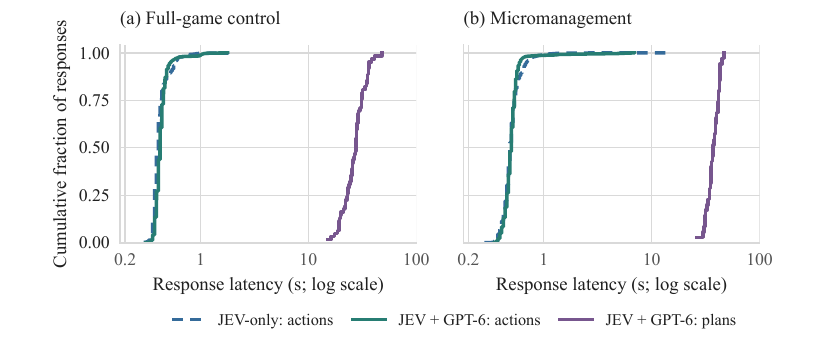}
\caption{Response-latency distributions by role and method. Both methods retain subsecond median JEV responses. GPT-6 provides less frequent, more expensive planning calls. The horizontal axis is logarithmic; each curve pools responses within its setting.}
\label{fig:latency}
\end{figure}

\paragraph{Frequent actions remain subsecond.}
Table~\ref{tab:latency} and Figure~\ref{fig:latency} report response distributions. Median JEV latency is 0.406 seconds for standalone macro control and 0.422 seconds with planning; corresponding micro medians are 0.485 and 0.500 seconds. In the combined system, macro and micro JEV 95th percentiles are 0.531 and 0.594 seconds. GPT-6 planning is less frequent but slower: median latency is 27.60 seconds in macro control and 37.17 seconds in micromanagement. These are different computational roles and inputs, so their ratio is not a same-task model speedup. The practical benefit is that most actions proceed through the faster decision layer.

\begin{table}[t]
\centering\small
\begin{tabular}{llrrrr}
\toprule
& & \multicolumn{2}{c}{JEV action responses} & \multicolumn{2}{c}{GPT-6 planning responses} \\
\cmidrule(lr){3-4}\cmidrule(lr){5-6}
Setting & Method & \(n\) & P50/P95 (s) & \(n\) & P50/P95 (s) \\
\midrule
Macro & JEV-only & 1,023 & 0.406/0.579 & 0 & -- \\
Macro & JEV + GPT-6 & 2,905 & 0.422/0.531 & 62 & 27.601/36.492 \\
Micro & JEV-only & 8,189 & 0.485/0.688 & 0 & -- \\
Micro & JEV + GPT-6 & 9,743 & 0.500/0.594 & 35 & 37.172/43.725 \\
 
\bottomrule
\end{tabular}
\captionbelowtable
\caption{Pooled latency over recorded responses with known usage. Macro JEV + GPT-6 statistics cover the four non-cheating wins; the Lv8 stress test is excluded. Micro statistics cover 105 episodes per method and 35 shared plans for the combined method.}
\label{tab:latency}
\end{table}

\begin{figure}[t]
\centering
\includegraphics[width=\linewidth]{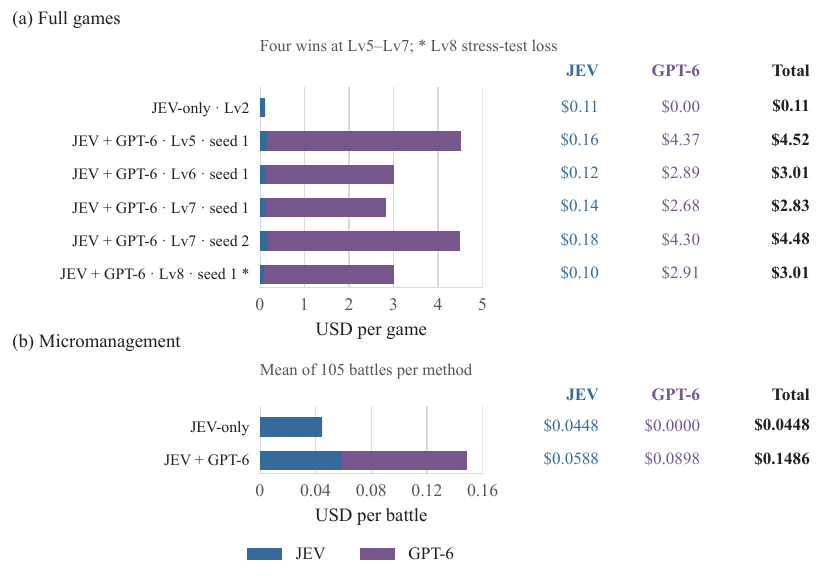}
\caption{Cost per game, with model components and totals. Top: each completed full-game experiment, including the separate Lv8 defeat. Bottom: mean battle cost over 105 episodes per method, amortizing each map's planning call across three episodes. Components and totals are rounded independently. Missing usage remains unpriced.}
\label{fig:costs}
\end{figure}

\paragraph{A complete game includes both models.}
Figure~\ref{fig:costs} shows the cost of each full game and the average micro episode, separating JEV, GPT-6, and their total. Across the four full-game wins, combined cost ranges from USD~2.83 to USD~4.52 and averages USD~3.7103. JEV accounts for USD~0.1499 per game and GPT-6 for USD~3.5604: planning contributes 96.0\% of the total. JEV's recorded macro usage costs approximately USD~0.21 per 1,000 responses. Cheap action selection therefore supports frequent interaction, while the planning schedule determines most of the game's model expenditure.

\begin{table}[t]
\centering\small
\setlength{\tabcolsep}{4pt}
\begin{tabular}{llrrrrr}
\toprule
& & & \multicolumn{3}{c}{USD per episode} & Wall time \\
\cmidrule(lr){4-6}
Setting & Method & Episodes & JEV & GPT-6 & Total & (s/episode) \\
\midrule
Macro & JEV-only & 1 & 0.1069 & 0.0000 & 0.1069 & 1239.8 \\
Macro & JEV + GPT-6 & 4 & 0.1499 & 3.5604 & 3.7103 & 828.6 \\
Micro & JEV-only & 105 & 0.0448 & 0.0000 & 0.0448 & 25.8 \\
Micro & JEV + GPT-6 & 105 & 0.0588 & 0.0898 & 0.1486 & 46.2 \\
 
\bottomrule
\end{tabular}
\captionbelowtable
\caption{Mean model cost and elapsed wall time per completed episode. Macro averages use the main full-game sample. Micro planner cost and setup time are amortized over three episodes per map. Costs are known-usage standard-API-equivalent estimates, not billing statements.}
\label{tab:costs}
\end{table}

For micromanagement, standalone JEV costs USD~0.0448 per episode. The combined method costs USD~0.1486, consisting of USD~0.0588 for JEV and USD~0.0898 for planning. Excluding the two development maps gives USD~0.0243 and USD~0.1168, respectively. The combined method thus spends more than JEV-only while producing better aggregate combat outcomes. Its advantage is improved capability with an inexpensive action layer, not a reduction in the total charge relative to the standalone controller.

\begin{figure}[t]
\centering
\includegraphics[width=\linewidth]{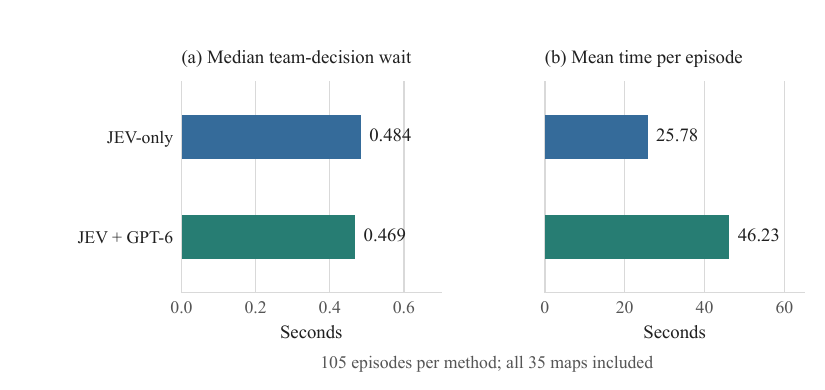}
\caption{Decision waiting time and total micromanagement runtime, shown directly for the two methods. (a) Median time to collect and decode all concurrent JEV batches for one team decision; planner calls are outside this timer. (b) Mean elapsed time per episode, calculated from each map's complete run divided by its three episodes, including planning, initialization, game interaction, and cleanup. All 35 maps and 105 episodes per method are included. Both panels use linear axes and seconds, with different ranges; each bar is labeled with its value.}
\label{fig:runtime}
\end{figure}

Mean micro wall time increases from 25.78 to 46.23 seconds per episode. This includes amortized planning, environment setup, stepping, and cleanup. Figure~\ref{fig:runtime} separates the wait for a complete team action from total elapsed time. Across 3,374 standalone and 4,089 combined-method decisions, median team waiting times are 0.484 and 0.469 seconds; their 95th percentiles are 1.537 and 1.954 seconds. Similar medians therefore do not imply similar total runtime. The combined method also runs longer battles on average (13.91 versus 11.48 game seconds), in addition to invoking the planner. The available timing records do not isolate how much each factor contributes to the total increase.

Macro games run in real time while inference overlaps game execution, and their durations also depend on when a match ends. Neither shorter successful games nor the difference between action and planning latency alone establishes a full-game speedup over prior systems.

\paragraph{Accounting basis.}
We price recorded token usage at the providers' standard rates on September 22, 2026 \citep{typesafe2026models,openai2026gpt6astra}. JEV charges for input; GPT-6 accounting separates cached from uncached input and includes output tokens. Reasoning output is a subset of output, not an additional charge. GPT-6 runs use Codex, so the reported amount is an equivalent API estimate. Calls without usage remain unpriced; Appendix~\ref{app:costs} gives token totals, missing-call counts, and per-game figures.

\section{Discussion and limitations}

The traces support a specific hypothesis about the standalone controller's weakness: local action selection does not reliably create and preserve a future objective. Saving for an expansion requires declining immediately feasible actions; coordinated attacks require maintaining posture despite short-term fluctuations. GPT-6 supplies these commitments, and the adapter makes them affect action selection. The logged progression from accepted goals to submitted actions supports this mechanism, without establishing what fraction of the performance gain is attributable to planning alone.

The comparison remains observational. The standalone and combined controllers also differ in candidate descriptions and execution handling, and the macro opponents are not matched. Four successful macro games and three micro episodes per map cannot establish broad reliability. A controlled next experiment should use the same executor and seeds with the planner enabled or disabled, followed by separate tests of resource reservation and posture persistence. The current data motivate the planning hypothesis without isolating its entire contribution.

The scope is also restricted: one macro race and matchup, one macro map, a custom starting-worker count, centralized micro observations, and separate macro and micro implementations. Low battle win counts remain a material weakness. Finally, inexpensive JEV inference does not make the combined system's planning free. Reducing unnecessary replanning or sharing plans more effectively is important because GPT-6 dominates observed full-game cost.

\section{Conclusion}

JEV-Star defeats Lv7, StarCraft II's strongest non-cheating built-in AI, through a combination of rapid structured decisions and persistent planning. Compared with the initial JEV-only controller, it also raises mean enemy elimination from 16.50\% to 37.69\% across 35 battle maps. Resource reservation and stable army objectives help explain the observed behavior. The four full-game wins retain a 0.422-second median JEV response and cost an estimated USD~3.71 per game. JEV provides inexpensive action selection, while GPT-6 supplies persistent strategic guidance and accounts for most model expenditure.

\begingroup
\small
\setlength{\bibsep}{2pt}
\bibliographystyle{plainnat}
\bibliography{references}

@article{vinyals2019alphastar,
  author = {Vinyals, Oriol and Babuschkin, Igor and Czarnecki, Wojciech M. and others},
  title = {Grandmaster level in {StarCraft II} using multi-agent reinforcement learning},
  journal = {Nature}, volume = {575}, pages = {350--354}, year = {2019},
  doi = {10.1038/s41586-019-1724-z},
  url = {https://www.nature.com/articles/s41586-019-1724-z}
}

@article{vinyals2017sc2le,
  author = {Vinyals, Oriol and Ewalds, Timo and Bartunov, Sergey and others},
  title = {{StarCraft II}: A New Challenge for Reinforcement Learning},
  journal = {arXiv preprint arXiv:1708.04782}, year = {2017},
  url = {https://arxiv.org/abs/1708.04782}
}

@article{samvelyan2019smac,
  author = {Samvelyan, Mikayel and Rashid, Tabish and de Witt, Christian Schr{\"o}der and Farquhar, Gregory and Nardelli, Nantas and Rudner, Tim G. J. and Hung, Chia-Man and Torr, Philip H. S. and Foerster, Jakob and Whiteson, Shimon},
  title = {The {StarCraft} Multi-Agent Challenge},
  journal = {arXiv preprint arXiv:1902.04043}, year = {2019},
  url = {https://arxiv.org/abs/1902.04043}
}

@article{deng2024smachard,
  author = {Deng, Yue and Yu, Yan and Ma, Weiyu and Wang, Zirui and Zhu, Wenhui and Zhao, Jian and Zhang, Yin},
  title = {{SMAC-Hard}: Enabling Mixed Opponent Strategy Script and Self-play on {SMAC}},
  journal = {arXiv preprint arXiv:2412.17707}, year = {2024},
  url = {https://arxiv.org/abs/2412.17707}
}

@article{ma2023textsc2,
  author = {Ma, Weiyu and Mi, Qirui and Zeng, Yongcheng and Yan, Xue and Wu, Yuqiao and Lin, Runji and Zhang, Haifeng and Wang, Jun},
  title = {Large Language Models Play {StarCraft II}: Benchmarks and A Chain of Summarization Approach},
  journal = {arXiv preprint arXiv:2312.11865}, year = {2023},
  note = {Version 3, June 2024}, url = {https://arxiv.org/abs/2312.11865}
}

@article{ahn2022saycan,
  author = {Ahn, Michael and Brohan, Anthony and Brown, Noah and others},
  title = {Do As I Can, Not As I Say: Grounding Language in Robotic Affordances},
  journal = {arXiv preprint arXiv:2204.01691}, year = {2022},
  url = {https://arxiv.org/abs/2204.01691}
}

@misc{typesafe2026jev,
  author = {Almeida, Diogo}, title = {Introducing System One Models \& Jev},
  year = {2026}, howpublished = {TypeSafe AI},
  url = {https://typesafe.ai/blog/introducing-system-one-models-and-jev},
  note = {Published September 15, 2026; accessed September 22, 2026}
}

@article{sun2018tstarbots,
  author = {Sun, Peng and Sun, Xinghai and Han, Lei and Xiong, Jiechao and Wang, Qing and Li, Bo and Zheng, Yang and Liu, Ji and Liu, Yongsheng and Liu, Han and Zhang, Tong},
  title = {{TStarBots: Defeating the Cheating Level Builtin AI in StarCraft II in the Full Game}},
  journal = {arXiv preprint arXiv:1809.07193}, year = {2018},
  url = {https://arxiv.org/abs/1809.07193}
}

@article{han2020tstarbotx,
  author = {Han, Lei and Xiong, Jiechao and Sun, Peng and Sun, Xinghai and Fang, Meng and Guo, Qingwei and Chen, Qiaobo and Shi, Tengfei and Yu, Hongsheng and Wu, Xipeng and Zhang, Zhengyou},
  title = {{TStarBot-X: An Open-Sourced and Comprehensive Study for Efficient League Training in StarCraft II Full Game}},
  journal = {arXiv preprint arXiv:2011.13729}, year = {2020},
  url = {https://arxiv.org/abs/2011.13729}
}

@article{wang2020scc,
  author = {Wang, Xiangjun and Song, Junxiao and Qi, Penghui and Peng, Peng and Tang, Zhenkun and Zhang, Wei and Li, Weimin and Pi, Xiongjun and He, Jujie and Gao, Chao and Long, Haitao and Yuan, Quan},
  title = {{SCC: an efficient deep reinforcement learning agent mastering the game of StarCraft II}},
  journal = {arXiv preprint arXiv:2012.13169}, year = {2020},
  url = {https://arxiv.org/abs/2012.13169}
}

@article{mathieu2023unplugged,
  author = {Mathieu, Micha{\"e}l and Ozair, Sherjil and Srinivasan, Srivatsan and Gulcehre, Caglar and others},
  title = {{AlphaStar Unplugged: Large-Scale Offline Reinforcement Learning}},
  journal = {arXiv preprint arXiv:2308.03526}, year = {2023},
  url = {https://arxiv.org/abs/2308.03526}
}

@article{ellis2022smacv2,
  author = {Ellis, Benjamin and Cook, Jonathan and Moalla, Skander and Samvelyan, Mikayel and Sun, Mingfei and Mahajan, Anuj and Foerster, Jakob N. and Whiteson, Shimon},
  title = {{SMACv2: An Improved Benchmark for Cooperative Multi-Agent Reinforcement Learning}},
  journal = {arXiv preprint arXiv:2212.07489}, year = {2022},
  url = {https://arxiv.org/abs/2212.07489}
}

@article{rashid2018qmix,
  author = {Rashid, Tabish and Samvelyan, Mikayel and de Witt, Christian Schroeder and Farquhar, Gregory and Foerster, Jakob and Whiteson, Shimon},
  title = {{QMIX: Monotonic Value Function Factorisation for Deep Multi-Agent Reinforcement Learning}},
  journal = {arXiv preprint arXiv:1803.11485}, year = {2018},
  url = {https://arxiv.org/abs/1803.11485}
}

@article{yu2021mappo,
  author = {Yu, Chao and Velu, Akash and Vinitsky, Eugene and Gao, Jiaxuan and Wang, Yu and Bayen, Alexandre and Wu, Yi},
  title = {{The Surprising Effectiveness of PPO in Cooperative, Multi-Agent Games}},
  journal = {arXiv preprint arXiv:2103.01955}, year = {2021},
  url = {https://arxiv.org/abs/2103.01955}
}

@article{li2024llmpysc2,
  author = {Li, Zongyuan and Ni, Yanan and Qi, Runnan and Jiang, Lumin and others},
  title = {{LLM-PySC2: Starcraft II learning environment for Large Language Models}},
  journal = {arXiv preprint arXiv:2411.05348}, year = {2024},
  url = {https://arxiv.org/abs/2411.05348}
}

@article{yao2022react,
  author = {Yao, Shunyu and Zhao, Jeffrey and Yu, Dian and Du, Nan and Shafran, Izhak and Narasimhan, Karthik and Cao, Yuan},
  title = {{ReAct: Synergizing Reasoning and Acting in Language Models}},
  journal = {arXiv preprint arXiv:2210.03629}, year = {2022},
  url = {https://arxiv.org/abs/2210.03629}
}

@article{shinn2023reflexion,
  author = {Shinn, Noah and Cassano, Federico and Berman, Edward and Gopinath, Ashwin and Narasimhan, Karthik and Yao, Shunyu},
  title = {{Reflexion: Language Agents with Verbal Reinforcement Learning}},
  journal = {arXiv preprint arXiv:2303.11366}, year = {2023},
  url = {https://arxiv.org/abs/2303.11366}
}

@article{wang2023voyager,
  author = {Wang, Guanzhi and Xie, Yuqi and Jiang, Yunfan and Mandlekar, Ajay and Xiao, Chaowei and Zhu, Yuke and Fan, Linxi and Anandkumar, Anima},
  title = {{Voyager: An Open-Ended Embodied Agent with Large Language Models}},
  journal = {arXiv preprint arXiv:2305.16291}, year = {2023},
  url = {https://arxiv.org/abs/2305.16291}
}

@article{liu2023agentbench,
  author = {Liu, Xiao and Yu, Hao and Zhang, Hanchen and Xu, Yifan and others},
  title = {{AgentBench: Evaluating LLMs as Agents}},
  journal = {arXiv preprint arXiv:2308.03688}, year = {2023},
  url = {https://arxiv.org/abs/2308.03688}
}

@article{chen2023frugalgpt,
  author = {Chen, Lingjiao and Zaharia, Matei and Zou, James},
  title = {{FrugalGPT: How to Use Large Language Models While Reducing Cost and Improving Performance}},
  journal = {arXiv preprint arXiv:2305.05176}, year = {2023},
  url = {https://arxiv.org/abs/2305.05176}
}

@article{ong2024routellm,
  author = {Ong, Isaac and Almahairi, Amjad and Wu, Vincent and Chiang, Wei-Lin and Wu, Tianhao and Gonzalez, Joseph E. and Kadous, M Waleed and Stoica, Ion},
  title = {{RouteLLM: Learning to Route LLMs with Preference Data}},
  journal = {arXiv preprint arXiv:2406.18665}, year = {2024},
  url = {https://arxiv.org/abs/2406.18665}
}

@article{kim2023llmcompiler,
  author = {Kim, Sehoon and Moon, Suhong and Tabrizi, Ryan and Lee, Nicholas and Mahoney, Michael W. and Keutzer, Kurt and Gholami, Amir},
  title = {{An LLM Compiler for Parallel Function Calling}},
  journal = {arXiv preprint arXiv:2312.04511}, year = {2023},
  url = {https://arxiv.org/abs/2312.04511}
}

@article{xu2023rewoo,
  author = {Xu, Binfeng and Peng, Zhiyuan and Lei, Bowen and Mukherjee, Subhabrata and Liu, Yuchen and Xu, Dongkuan},
  title = {{ReWOO: Decoupling Reasoning from Observations for Efficient Augmented Language Models}},
  journal = {arXiv preprint arXiv:2305.18323}, year = {2023},
  url = {https://arxiv.org/abs/2305.18323}
}

@article{zhou2023webarena,
  author = {Zhou, Shuyan and Xu, Frank F. and Zhu, Hao and Zhou, Xuhui and Lo, Robert and Sridhar, Abishek and Cheng, Xianyi and Ou, Tianyue and Bisk, Yonatan and Fried, Daniel and Alon, Uri and Neubig, Graham},
  title = {{WebArena: A Realistic Web Environment for Building Autonomous Agents}},
  journal = {arXiv preprint arXiv:2307.13854}, year = {2023},
  url = {https://arxiv.org/abs/2307.13854}
}

@misc{openai2026gpt6astra,
  author = {{OpenAI}}, title = {{gpt-6-astra} model documentation and pricing},
  year = {2026}, url = {https://developers.openai.com/api/docs/models/gpt-6-astra},
  note = {Standard API rates; accessed September 22, 2026}
}

@misc{typesafe2026models,
  author = {{TypeSafe AI}}, title = {Models and pricing}, year = {2026},
  url = {https://docs.typesafe.ai/models}, note = {Accessed September 22, 2026}
}

@misc{blizzard2026sc2protocol,
  author = {{Blizzard Entertainment}},
  title = {{StarCraft II} API protocol: built-in opponent difficulty levels},
  year = {2026},
  url = {https://github.com/Blizzard/s2client-proto/blob/master/s2clientprotocol/sc2api.proto},
  note = {Difficulty enumeration; accessed September 22, 2026}
}
\endgroup

\appendix
\clearpage
\section{Replay frames and decision traces}
\label{app:replays}

\subsection{A paired micromanagement example}

\begin{figure}[H]
\centering
\includegraphics[width=\linewidth]{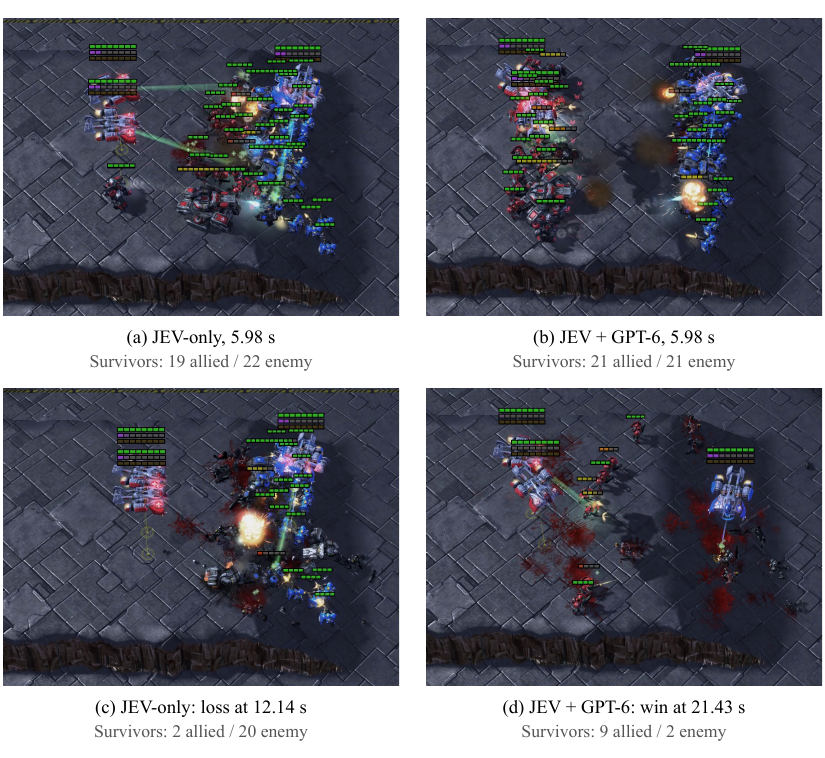}
\caption{Replay frames from \texttt{mmmt}, episode 1. The top row shows both methods at the same game time, 5.98 seconds. The bottom row shows each episode's terminal state, at different times. Allied units are red and opponents blue. Counts include surviving non-attacking support units, so a benchmark victory can leave enemy Medivacs alive. The same camera settings and fixed image crop are used for both columns.}
\label{fig:microreplay}
\end{figure}

Figure~\ref{fig:microreplay} compares the first \texttt{mmmt} episode for each method. Both start with 22 allied and 22 enemy units. JEV-only loses at 12.14 seconds with two allied units remaining; JEV + GPT-6 wins at 21.43 seconds with nine. The winning controller's recorded guidance calls for an infantry screen, supported tanks, and healing of wounded units. This is one illustrative pair: the complete map result is zero versus two wins out of three, reported alongside all other maps.

\clearpage
\subsection{Battle progression in the paired replay}

\begin{figure}[h]
\centering
\includegraphics[width=\linewidth]{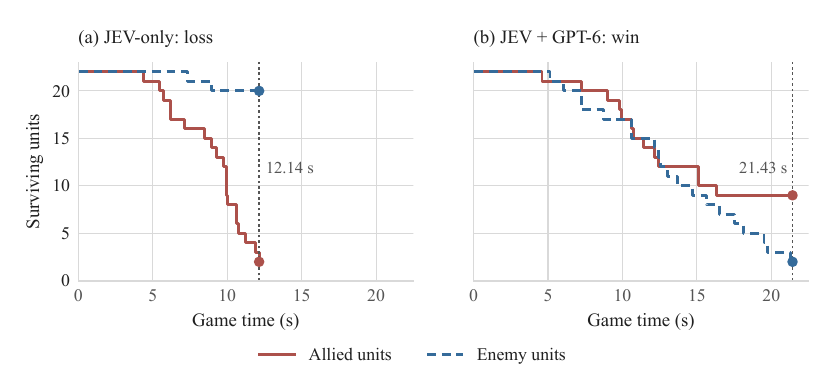}
\caption{Surviving units in the paired \texttt{mmmt} replays. The panels use identical axes and include all units, including non-attacking support units. Counts come from 272 and 480 decoded replay frames, respectively, and agree with the recorded battle states. Vertical dotted lines mark the terminal times, 12.14 and 21.43 seconds; the shorter episode is not extrapolated to the longer duration. This is one episode per controller, not a win-rate estimate.}
\label{fig:attrition}
\end{figure}

Figure~\ref{fig:attrition} reconstructs surviving-unit counts from the same two replays. Both start with 22 units on each side. The standalone controller ends with two allied and 20 enemy units; the combined controller ends with nine allied and two enemy units. Each curve ends at its own episode boundary. These trajectories describe this pair of battles, while the aggregate evaluation remains the full 35-map comparison.

\clearpage
\subsection{Full-game decision trace}

This excerpt projects recorded fields from the Lv7 seed-1 log; times and response latency are rounded. Action identifier 21 denotes building a Nexus. The observation, JEV response, command, and game result are separate events.

\begin{lstlisting}
[01:34.286] plan_accepted
  base_target: 2; worker_target: 30
  guidance (exact sentence):
    "Start the natural around 19 workers before filling army ceilings."

[02:56.250] request
  resource.mineral: 400; building.nexus_count: 1

[02:56.786] response
  answer.choice: "21"  (BUILD NEXUS)
  answer.confidence: 0.83; latency_ms: 391

[02:56.786] action
  description: "BUILD NEXUS"
  orders_submitted: 1
  orders[0].ability: "PROTOSSBUILD_NEXUS"

[07:32.321] plan_accepted
  army_posture: "attack"
  attack_min_army: 44; retreat_below_army: 26
  min_posture_seconds: 30

[07:33.214] action
  description: "MULTI-ATTACK"
  orders_submitted: 24

[12:47.679] end
  result: "Victory"
\end{lstlisting}

\subsection{Micromanagement plan and outcome}

These sentences are quoted from the accepted tactical plan for \texttt{mmmt}; the outcome fields come from its first episode. The plan is reused across the map's three episodes. Its requested behavior is distinct from whether each unit follows that behavior successfully.

\begin{lstlisting}
plan_accepted | map: "mmmt"
opening (exact sentences):
  "Advance together east through the central corridor, y in [11,21), toward the raised center."
  "Keep infantry screening tanks and Medivacs behind infantry."
movement_and_cooldowns (exact sentences):
  "Ready, in-range shooters normally fire."
  "Withdraw a badly wounded unit taking recent damage toward healing while healthy allies maintain fire."
episode_end
  episode: 1; status: "win"
  final_game_loop: 480; game_seconds: 21.429
  final.allies_alive: 9; final.enemies_alive: 2
\end{lstlisting}

\section{All recorded micromanagement victories}
\label{app:allwins}

These galleries cover all ten recorded wins across six maps. Each row pairs combat at 40\% of the episode duration with its terminal state. Clocks restart at the episode boundary. Camera distance is shared; crops are fixed within maps, with a wider view for \texttt{2c\_vs\_64zg}. Terminal counts match the episode summaries and include non-attacking support units.

\begin{figure}[H]
\centering
\includegraphics[width=\linewidth]{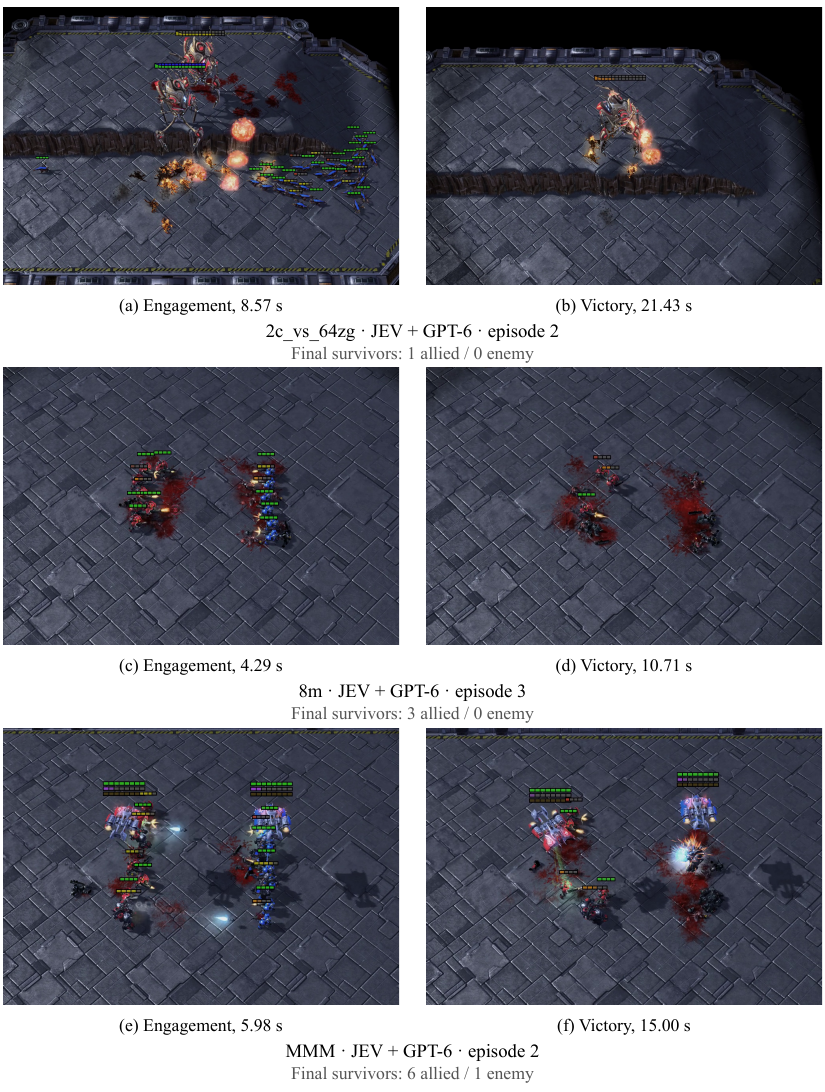}
\caption{Three JEV + GPT-6 victories on distinct maps: \texttt{2c\_vs\_64zg}, episode 2; \texttt{8m}, episode 3; and \texttt{MMM}, episode 2. Left: combat. Right: terminal state. The final allied/enemy counts are 1/0, 3/0, and 6/1; the remaining enemy in \texttt{MMM} is a non-attacking support unit. Each row identifies its map, controller, and episode.}
\label{fig:winsranged}
\end{figure}

\clearpage
\begin{figure}[H]
\centering
\includegraphics[width=\linewidth]{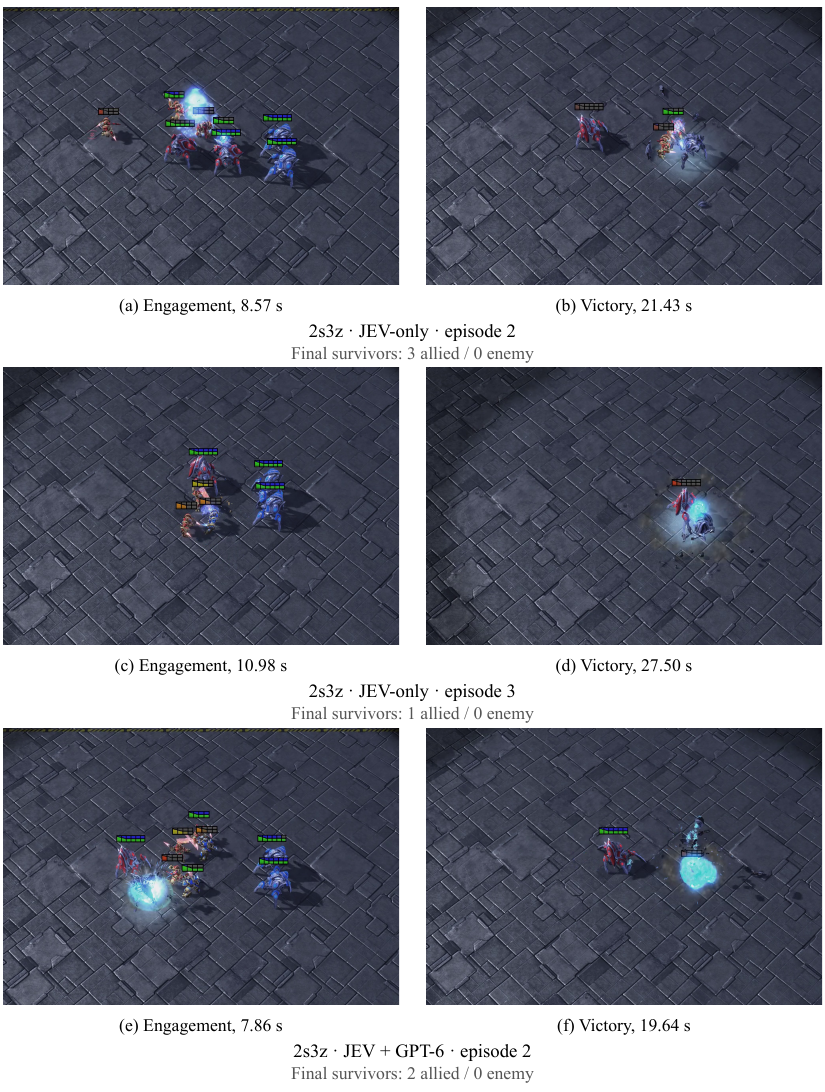}
\caption{All three \texttt{2s3z} victories: JEV-only episodes 2 and 3, and JEV + GPT-6 episode 2. Each row shows engagement and terminal frames from the same episode. Final allied/enemy counts are 3/0, 1/0, and 2/0. The standalone controller wins twice on this map while the combined controller wins once; the gallery retains this unfavorable comparison.}
\label{fig:wins2s3z}
\end{figure}

\clearpage
\begin{figure}[H]
\centering
\includegraphics[width=\linewidth]{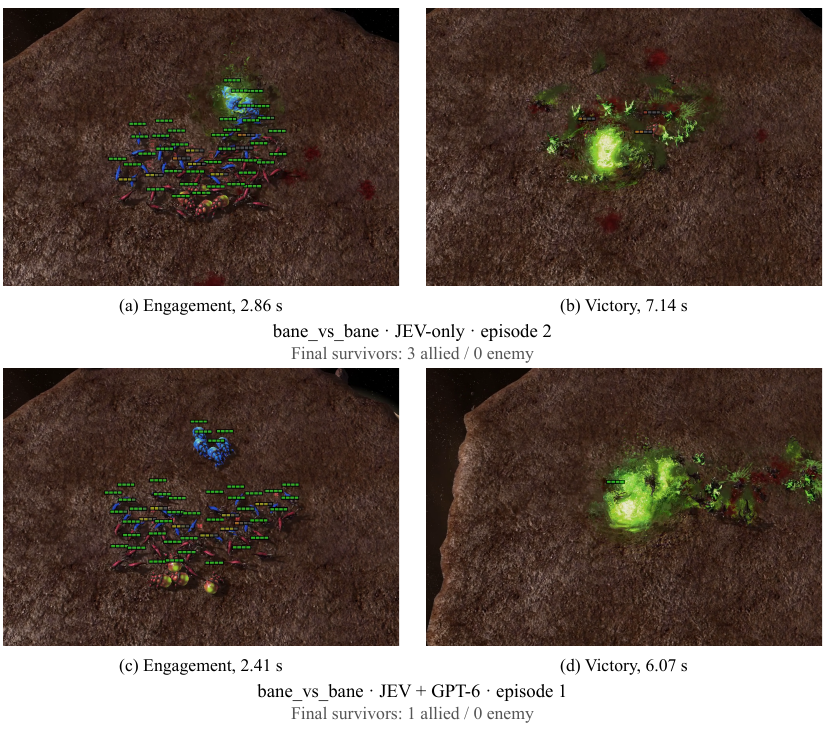}
\caption{Both \texttt{bane\_vs\_bane} victories: JEV-only episode 2 and JEV + GPT-6 episode 1. Engagement frames precede the terminal states at 7.14 and 6.07 episode seconds. Three and one allied units survive, respectively, with no enemies remaining. Earlier episodes contained in the replay files are excluded from the displayed episode clocks.}
\label{fig:winsbane}
\end{figure}

\clearpage
\begin{figure}[H]
\centering
\includegraphics[width=\linewidth]{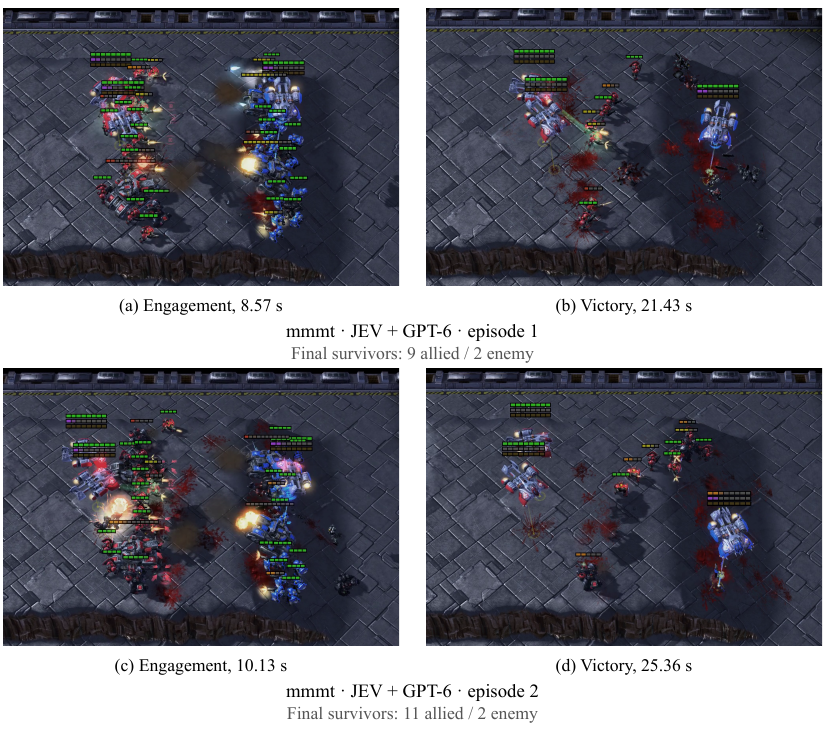}
\caption{Both JEV + GPT-6 victories on \texttt{mmmt}, episodes 1 and 2. Final allied/enemy counts are 9/2 and 11/2; the two surviving enemies are non-attacking support units. The second episode lasts 25.36 seconds, independently of the preceding battle's clock. The first row corresponds to the combined-system episode in the paired comparison of Figure~\ref{fig:microreplay}.}
\label{fig:winsmmmt}
\end{figure}

\clearpage
\section{Opening construction and conditional unit choices}
\label{app:macrochoices}

Table~\ref{tab:opening} lists every submitted construction order in the first four game minutes of the standalone run and the Lv7 seed-1 win, including repeated Pylons and Gateways. Probe production and research are excluded from this construction sequence and analyzed separately in Section~\ref{sec:macrochoices}. The worker column records the original request observation, which can precede the asynchronous order. These games start with eight workers.

\begin{table}[H]
\centering\small
\setlength{\tabcolsep}{5pt}
\begin{tabular}{rrlrrl}
\toprule
\multicolumn{3}{c}{JEV-only, Lv2} & \multicolumn{3}{c}{JEV + GPT-6, Lv7 seed 1} \\
\cmidrule(lr){1-3}\cmidrule(lr){4-6}
Time & Workers & Construction & Time & Workers & Construction \\
\midrule
00:29 & 10 & Pylon & 00:30 & 10 & Pylon \\
00:41 & 11 & Assimilator & 00:45 & 11 & Assimilator \\
01:06 & 13 & Gateway & 01:11 & 13 & Gateway \\
01:29 & 14 & Assimilator & 01:39 & 16 & Pylon \\
01:41 & 15 & Forge & 02:01 & 17 & Cybernetics Core \\
02:06 & 17 & Pylon & 02:31 & 20 & Pylon \\
02:41 & 18 & Cybernetics Core & 02:57 & 22 & Nexus \\
02:54 & 19 & Photon Cannon & 03:25 & 24 & Gateway \\
03:15 & 19 & Pylon & 03:28 & 25 & Assimilator \\
03:30 & 20 & Shield Battery & 03:34 & 25 & Shield Battery \\
03:49 & 21 & Twilight Council & 03:44 & 26 & Pylon \\
 
\bottomrule
\end{tabular}
\captionbelowtable
\caption{Opening build order through 4:00 game time. All rows are submitted construction commands; times do not denote completion. Repeated structures remain in the sequence.}
\label{tab:opening}
\end{table}

Table~\ref{tab:unitchoices} conditions unit selection on the options actually offered to JEV. An exposure is one processed decision whose original candidate set includes that unit; several units can be offered together. Consequently, denominators differ by row and cannot be added as disjoint opportunities. The combined column pools the four main wins. Differences in resource reservation, production priorities, prerequisites, and idle producers affect these candidate sets, so the table describes the complete controller rather than measuring preference under identical alternatives.

\begin{table}[H]
\centering\small
\begin{tabular}{lrr}
\toprule
Production choice & JEV-only: selected / offered & JEV + GPT-6: selected / offered \\
\midrule
Zealot & 29 / 95 & 156 / 361 \\
Stalker & 8 / 42 & 149 / 233 \\
Adept & 6 / 85 & 30 / 59 \\
Sentry & 17 / 58 & -- \\
Immortal & 0 / 3 & 26 / 34 \\
Colossus & -- & 5 / 5 \\
High Templar & -- & 9 / 10 \\
Observer & 8 / 14 & 6 / 10 \\
 
\bottomrule
\end{tabular}
\captionbelowtable
\caption{Conditional non-worker unit choices in the main full-game samples. A dash means the candidate was never offered. Each selected action in these rows submits one production order; this does not establish that every unit finishes production.}
\label{tab:unitchoices}
\end{table}

The action profile contains 1,020 standalone and 2,871 combined-system action events. Three standalone and 34 combined-system response records have no corresponding action event and are excluded from that profile. The source-linked analysis retains all per-game counts, the offered choices, order submission times, and observed research completion times. The independent Lv8 game remains outside the main action and technology comparison.

\clearpage
\section{Per-map micromanagement results}
\label{app:maps}

Each map has three episodes per method. The table reports wins and mean enemy elimination; daggers identify the two local development maps. All 35 maps are included in the main 105-episode result.

\begin{table}[H]
\centering\small
\begin{tabular}{lrrrr}
\toprule
& \multicolumn{2}{c}{Wins / 3} & \multicolumn{2}{c}{Enemy elimination (\%)} \\
\cmidrule(lr){2-3}\cmidrule(lr){4-5}
Map & JEV-only & JEV + GPT-6 & JEV-only & JEV + GPT-6 \\
\midrule

3m & 0 & 0 & 0.0 & 66.7 \\
8m & 0 & 1 & 16.7 & 70.8 \\
25m & 0 & 0 & 1.3 & 54.7 \\
5m\_vs\_6m & 0 & 0 & 5.6 & 33.3 \\
8m\_vs\_9m & 0 & 0 & 0.0 & 40.7 \\
10m\_vs\_11m & 0 & 0 & 9.1 & 48.5 \\
27m\_vs\_30m & 0 & 0 & 2.2 & 27.8 \\
MMM & 0 & 1 & 10.0 & 73.3 \\
MMM2 & 0 & 0 & 2.8 & 30.6 \\
2s3z & 2 & 1 & 86.7 & 66.7 \\
3s5z & 0 & 0 & 41.7 & 62.5 \\
3s5z\_vs\_3s6z & 0 & 0 & 25.9 & 25.9 \\
3s\_vs\_3z & 0 & 0 & 11.1 & 33.3 \\
3s\_vs\_4z & 0 & 0 & 0.0 & 25.0 \\
3s\_vs\_5z & 0 & 0 & 0.0 & 20.0 \\
1c3s5z & 0 & 0 & 40.7 & 55.6 \\
2m\_vs\_1z & 0 & 0 & 0.0 & 0.0 \\
corridor & 0 & 0 & 50.0 & 61.1 \\
6h\_vs\_8z & 0 & 0 & 4.2 & 37.5 \\
2s\_vs\_1sc & 0 & 0 & 0.0 & 0.0 \\
so\_many\_baneling & 0 & 0 & 47.9 & 59.4 \\
bane\_vs\_bane & 1 & 1 & 100.0 & 77.8 \\
2c\_vs\_64zg & 0 & 1 & 22.9 & 75.5 \\
2vr\_vs\_3sc & 0 & 0 & 0.0 & 0.0 \\
3hl\_vs\_24zl & 0 & 0 & 0.0 & 15.3 \\
3rp\_vs\_5zl & 0 & 0 & 0.0 & 0.0 \\
3rp\_vs\_24zl & 0 & 0 & 0.0 & 5.6 \\
7q\_vs\_2bc & 0 & 0 & 0.0 & 0.0 \\
3st\_vs\_5zl & 0 & 0 & 6.7 & 20.0 \\
6m\_vs\_10m & 0 & 0 & 0.0 & 20.0 \\
mmmt\_vs\_zspi & 0 & 0 & 16.7 & 47.9 \\
unit\_test$^\dagger$ & 0 & 0 & 0.0 & 0.0 \\
mmmt & 0 & 2 & 3.0 & 87.9 \\
mmmt\_vs\_zhb & 0 & 0 & 8.0 & 49.3 \\
pvt\_large$^\dagger$ & 0 & 0 & 64.4 & 26.7 \\
 
\bottomrule
\end{tabular}
\captionbelowtable
\caption{Complete comparison of the two methods on the same maps.}
\label{tab:allmaps}
\end{table}

\clearpage
\section{Cost accounting and reproducibility}
\label{app:costs}

Table~\ref{tab:individualcosts} lists the charges underlying Figure~\ref{fig:costs}. Each total is the sum of the two model components. The main combined macro average includes only the four Lv5--Lv7 games; the Lv8 stress test remains a separate completed result.

\begin{table}[h]
\centering\small
\begin{tabular}{llcrrr}
\toprule
Method & Difficulty & Seed & JEV (USD) & GPT-6 (USD) & Total (USD) \\
\midrule
JEV-only & Lv2 & 1 & 0.1069 & 0.0000 & 0.1069 \\
JEV + GPT-6 & Lv5 & 1 & 0.1551 & 4.3683 & 4.5233 \\
JEV + GPT-6 & Lv6 & 1 & 0.1169 & 2.8885 & 3.0053 \\
JEV + GPT-6 & Lv7 & 1 & 0.1449 & 2.6842 & 2.8291 \\
JEV + GPT-6 & Lv7 & 2 & 0.1829 & 4.3006 & 4.4834 \\
JEV + GPT-6 & Lv8 & 1 & 0.1029 & 2.9068 & 3.0098 \\
 
\bottomrule
\end{tabular}
\captionbelowtable
\caption{Individual full-game costs estimated from known token usage.}
\label{tab:individualcosts}
\end{table}

The rates are USD~0.042 per million JEV input tokens with no output charge, and USD~10, 1, and 50 per million GPT-6 uncached input, cached input, and output tokens, respectively \citep{typesafe2026models,openai2026gpt6astra}. Cache-write input, if present, uses USD~12.50 per million. Requests exceeding 272,000 input tokens use the published long-context multipliers of two for input and 1.5 for output. We apply rates to each response before aggregation. These rates serve only to reconstruct game expenditure; the main comparison uses actual workload totals.

\begin{table}[h]
\centering\scriptsize
\setlength{\tabcolsep}{3.5pt}
\begin{tabular}{llrrrrr}
\toprule
Setting & Method & JEV input & GPT-6 input & Cached input & GPT-6 output & Unpriced J/G \\
\midrule
Macro & JEV-only & 2,544,622 & 0 & 0 & 0 & 1/0 \\
Macro & JEV + GPT-6 & 14,278,950 & 1,437,426 & 166,400 & 27,297 & 2/4 \\
Micro & JEV-only & 111,915,677 & 0 & 0 & 0 & 0/0 \\
Micro & JEV + GPT-6 & 147,109,930 & 847,569 & 30,720 & 24,498 & 16/0 \\
 
\bottomrule
\end{tabular}
\captionbelowtable
\caption{Recorded token totals for the main samples. Cached input is included in GPT-6 input. Unpriced J/G counts identify JEV/GPT-6 attempts without recorded usage; their unknown cost is omitted from the known-usage estimate.}
\label{tab:tokens}
\end{table}

The Lv8 stress test additionally contains one GPT-6 attempt without usage. A canceled planning attempt before the selected micro run also lacks usage and falls outside the completed-episode totals. Neither is treated as free. Micro planning costs are divided by three because each map's plan is reused for three episodes; they would be higher per game without reuse.

The accompanying analysis files contain 210 battle-episode records, response-level latency and usage, full-game action and state trajectories, and accepted plans. A selection manifest links these records to frozen source summaries and hashed event-log prefixes. Later appends to logs do not alter the selected data. Request latency is measured over responses with known usage; wall time comes from complete run summaries and includes setup and cleanup.

Reproduction requires the stated StarCraft II build, map files, shared-vision and ability settings, and the recorded action adapters. The initial macro catalog has 72 commands and the combined controller has 73; candidate and execution improvements are part of the system comparison. No model weights are trained.

\end{document}